\documentclass[11pt,twoside]{article}
\usepackage[pdftex]{graphicx}
\usepackage{amsmath}
\usepackage{xcolor}
\usepackage{amssymb}
\usepackage{cite}

\begin{document}
\newcommand{\pst}{\hspace*{1.5em}}

\newcommand{\rigmark}{\em Journal of Russian Laser Research}
\newcommand{\lemark}{\em Volume 30, Number 5, 2009}

\newcommand{\be}{\begin{equation}}
\newcommand{\ee}{\end{equation}}
\newcommand{\bm}{\boldmath}
\newcommand{\ds}{\displaystyle}
\newcommand{\bea}{\begin{eqnarray}}
\newcommand{\eea}{\end{eqnarray}}
\newcommand{\ba}{\begin{array}}
\newcommand{\ea}{\end{array}}
\newcommand{\arcsinh}{\mathop{\rm arcsinh}\nolimits}
\newcommand{\arctanh}{\mathop{\rm arctanh}\nolimits}
\newcommand{\bc}{\begin{center}}
\newcommand{\ec}{\end{center}}

\thispagestyle{plain}

\label{sh}


\begin{center} {\Large \bf
\begin{tabular}{c}
DISCRETIZATION OF THE DENSITY MATRIX AS
\\[-1mm]
A NONLINEAR POSITIVE MAP AND ENTANGLEMENT
\end{tabular}
 } \end{center}

\bigskip

\bigskip

\begin{center} {\bf
Julio A.  L\'opez-Sald\'ivar,$^{1*}$,  Armando Figueroa,$^1$, Octavio Casta\~nos,$^1$, Ram\'on L\'opez--Pe\~na$^1$, \smallskip  Margarita A. Man'ko$^{2}$ and Vladimir I. Man'ko$^{2,3}$
}\end{center}

\medskip

\begin{center}
{\it
$^1$Instituto de Ciencias Nucleares, Universidad Nacional Aut\'onoma de M\'exico, Apartado Postal 70-543, 04510 M\'exico DF,   Mexico 
\smallskip

$^2$Lebedev Physical Institute, Leninskii Prospect 53,
Moscow 119991, Russia 
\smallskip

$^3$Moscow Institute of Physics and Technology (State University) Dolgoprudnyi, Moscow Region 141700, Russia

$^*$Corresponding author e-mail:~~~julio.lopez~@~nucleares.unam.mx\\}
\end{center}

\begin{abstract}\noindent
The discretization of the density matrix is proposed as a nonlinear
positive map for systems with continuous variables. This procedure
is used to calculate the entanglement between two modes through
different criteria, such as Tsallis entropy, von Neumann entropy and
linear entropy and the logarithmic negativity. As an example, we
study the dynamics of entanglement for the two-mode squeezed vacuum
state in the parametric amplifier and show good agreement with the
analytic results. The loss of information on the system state due to
the discretization of the density matrix is also addressed.
\end{abstract}

\medskip


\section{Introduction}
\pst
The pure state of a quantum system is identified with the wave
function~\cite{Schrod26}, while its mixed state is identified with
the density matrix~\cite{Landau27,vonNeumann27,vonNeumann32}. The
time evolution of the quantum-system state provides the
transformation (map) of the initial-state density matrix $\rho(0)$
onto the density matrix at time $t$, i.e., $\rho(0)\to\rho(t)$. The
density matrix has nonnegative eigenvalues only, and such
transformations are called positive maps. Among the positive maps,
there are linear positive maps and nonlinear ones. The set of linear
positive maps contains a subset of completely positive maps. The
properties of the maps were considered in
\cite{Choi,Stinespring,SudarMathewsRaoPRA61}, and nonlinear maps
were discussed in \cite{PuzkoEPL,VentriPLA}.

A positive map is a transformation of a density matrix into a
density matrix, i.e., this kind of operation preserves the trace of
the density matrix, its hermiticity and positivity. Any quantum
operation can be expressed as a positive map. In particular, the
study of completely positive maps have been of great interest in the
quantum information theory; it provides the result that the
transmission of classical or quantum information can be represented
by a set of operators defining a completely positive map, also
called the quantum channel.

There exist different capacities of a system to transmit
information: the classical capacity~\cite{clas_Hausladen}, the
quantum capacity~\cite{cuan_Lloyd}, and the entangled assisted
capacity~\cite{enr_bennet,enr_bennet02}. Some of these capacities
can be calculated through the three von Neumann entropies, such as
the entropy of the input system state $\rho$, the entropy associated
to the map $\Phi (\rho)$, and the entropy exchange $(\Phi \otimes
\textrm{Id})\vert \psi \rangle \langle \psi \vert$, where $\vert
\psi \rangle$ denotes the state which, under purification over an
ancillary system $R$, gives the input $\rho$ ($\rho={\rm Tr}_R (\vert
\psi\rangle \langle \psi \vert)$).

The constructions of two linear positive maps of a qudit system are
used to implement new entropic inequalities in  composite and
non-composite
systems~\cite{manko1,chernega1,manko2,chernega2,manko_ent,manko3}.
This method transforms the density $N$$\times$$N$-matrix with $N=nm$
into two density matrices $\rho^{(1)}$ and $\rho^{(2)}$ with
dimensions $m$$\times$$m$ and $n$$\times$$n$, respectively.

It is known that non-unitary transformation, such as the generic
time evolution of the density matrix can be expressed in terms of a
completely positive map, in view of the Kraus
decomposition~\cite{kraus}. For finite-dimensional density matrices,
such maps were discussed in \cite{SudarMathewsRaoPRA61}. These
linear maps have also been used to construct the time evolution of
the initial state for any master equation which is local in time,
whether Markovian or non-Markovian of
Gorini--Kossakowski--Sudarshan~\cite{CKS} and
Lindblad~\cite{Lindblad} form or not in~\cite{andersson}. In
\cite{manko_maps}, a positive map that changes a mixed state into
the pure state was introduced. Pseudo-positive linear maps were
considered in \cite{CrusPS,TraskunovJRLR}.

In this study, we propose the discretization of the density matrix
of continuous variables as a nonlinear map that preserves the
quantum properties of the system. We show that the discrete form of
the density matrix can be used to calculate the entanglement
measures and other observables numerically. Implementing another map
which reduces the discrete $N$$\times$$N$ density matrix to an
$n$$\times$$n$ density matrix (with $n<N$), we study the loss of
information due to the discretization procedure.

This paper is organized as follows.

In section 2, we define the discretization of the density matrix of
continuous variables, establishing the conditions that should be
satisfied to correspond to the positive map. Also in this section
the positive map of a discrete $N$$\times$$N$ density matrix into a
cut discrete $n$$\times$$n$ density matrix ($n<N$) is discussed in
order to address the loss of information on the system. The
entanglement properties of the squeezed vacuum state in the
parametric amplifier are obtained in section 3 as an example of the
application of the discretization procedure. The entanglement is
calculated through different quantities, such as Tsallis entropy,
von Neumann entropy and linear entropy of the cut density matrices
and the logarithmic negativity for the two-mode partial transpose.
The calculation of the covariance matrix is discussed in section 4.
Numerical results are compared with the analytical expressions in
sections 3 and 4. Finally, some conclusions are presented.

\section{Discretization as a nonlinear positive map of the density matrices}

In this section, we consider for two-mode field the procedure of
discretization of the density matrix $\rho^{(12)}
(x,y;x',y')=\langle x,y \vert
\hat{\rho}^{(12)} \vert x', y' \rangle$, which satisfies the
normalization condition
\begin{equation}
\intop_{-\infty}^{\infty} \intop_{-\infty}^{\infty} \rho^{(12)} (x,y;x,y) \, dx \, dy=1.
\label{norm_cond}
\end{equation}
To obtain a discrete form, first, we define the discrete numbers of
the axes $x_i = i \Delta x$, $y_j= j \Delta y$, $x'_k=k \Delta x$
and $y'_l=l \Delta y$ with $i,j,k,l=0,\pm 1, \pm 2, \ldots, \pm M$.
In addition, the steps $\Delta x$, $\Delta y$, $\Delta x'$ and
$\Delta y'$ are such that the numerical convergence of the
normalization condition~(\ref{norm_cond}) is ensured. This
discretization provides the density matrix
\begin{equation}
\rho^{(12)} (x_i, y_j ; x'_k, y'_l)=\rho^{(12)}_{ij,kl} \ ,
\label{d1}
\end{equation}
and this relation constitutes a map of infinite-dimensional Hilbert
space $\mathcal{H}$ onto the finite-dimensional Hilbert space
$\mathcal{H}_d$. In order to define correctly the conjugate
transpose matrix $\rho^{(12)}_{ij,kl}$ and the normalization
condition, the steps in the variables $x$ and $x'$ should be $\Delta
x=\Delta x'$ and also $\Delta y=\Delta y'$. Then the discrete
normalization reads
\begin{equation}
\intop_{-\infty}^{\infty} \intop_{-\infty}^{\infty} \rho^{(12)} (x,y;x,y) \, dx \, dy
= \sum_{i,j=-M}^M \rho^{(12)}_{ij,ij} \Delta x \Delta y=1 \, ,
\end{equation}
and the conjugate transpose matrix is $
\rho^{(12)\,\dagger}_{ij,kl}=\rho^{(12)\,*}_{kl,ij}\, .
$

One can see that the obtained discrete density matrix
$\rho^{(12)}_{ij,kl}$ is a result of action by the nonlinear
positive map of the continuous matrix $\rho^{(12)} (x,y;x',y')$
satisfying the same properties as a standard density matrix, i.e.,
is a hermitian, positive, semi-definite one with $\textrm{Tr}
\rho^{(12)}=1$. Thus, information on the initial continuous matrix can be
obtained through information on the discrete density matrix obtained
by means of the positive map.

The partial density matrices for the system can be calculated as
\begin{equation}
\rho^{(1)}_{ij}=\sum_{k=-M}^{M} \rho^{(12)}_{i k,j k} \Delta y\, , \quad
\rho^{(2)}_{ij}=\sum_{k=-M}^{M} \rho^{(12)}_{k i,k j} \Delta x \, ,
\end{equation}
where as in (\ref{d1}) we now use
\[
\rho^{(k)}_{ij}=\rho^{(k)}(z,z^\prime) \, ,
\]
with $k=1,2$; for the first mode $z=x$, while for the second one
$z=y$. One has the following discrete normalization condition:
\[
\intop_{-\infty}^{\infty}\rho^{(k)}(z,z)dz \approx\sum_{i=-M}^{M}
\rho^{(k)}_{ii}\Delta z=\textrm{Tr}\,\rho^{(k)}=\sum_{i}R^{(k)}_{ii}=1 \,
.
\]

We choose the interval $dz$ in the sum form of the integral
providing that $\rho^{(k)}_{ii}\Delta z=R^{(k)}_{ii}$. This discrete
form $\rho^{(k)}_{ij}$ is a nonlinear positive map of the reduced
density matrix for the system, which can be used to calculate all
the observables, such as, for example, the entanglement
entropies~\cite{julio}.

We assume $\rho^{(k)}_{i j}<\varepsilon$ for $i,j>(2M+1)$, where
$\varepsilon$ is arbitrarily small. Then the infinite matrix
$R^{(k)}_{ij}$ has the approximation
\[
R^{(k)}_{ij}=\left(\begin{array}{cccccc}
0 & 0 & 0 & \cdots & 0 & 0\\
0 & R_{-M,-M} & R_{-M,-M+1} & \cdots & R_{-M,M} & 0\\
0 & R_{-M+1,-M} & R_{-M+1,-M+1} & \cdots & R_{-M+1,M} & 0\\
\vdots & \vdots & \vdots & \vdots & \vdots & \vdots\\
0 & R_{M,-M} & R_{M,-M+1} & \cdots & R_{M,M} & 0\\
0 & 0 & 0 & \cdots & 0 & 0
\end{array}\right),
\]
where $R^{(k)}_{ij}$ has an {\it ocean of zeros}. For all the
reduced density matrices $\rho^{(k)}$, such that
$\textrm{Tr}\,\rho^{(k)}=1$, the discretization procedure provides
the infinite matrix $R^{(k)}_{ij}$, which has the form of an {\it
island} finite $N$$\times$$N$ matrix ($N=2M+1$) floating in the infinity ocean
of zero matrix elements.  Thus we have mapped the infinite matrix
onto a finite $N$$\times$$N$ density matrix.

To give an example of such maps, we discuss a nonlinear positive
map, which transforms an $N$$\times$$N$ finite density matrix onto a
$n$$\times$$n$ density matrix with $n<N$. The initial $N$$\times$$N$
density matrix $\rho$ for mode~1 has the matrix elements
\[
\rho^{(1)}_{ij}=\left\langle i\right| \rho^{(1)}\left|j\right\rangle.\]
Let us construct the matrix
\[
R^{(1)}_{o\, jk}=\left\langle j\right|{R}^{(1)}\left|k\right\rangle
,
\]
following the rule that all the matrix elements $R^{(1)}_{2p,2s}$
with $p,s=1,2,\ldots,n$ are equal to zero. Let the matrix elements
$R^{(1)}_{o\,ps}=R^{(1)}_{ij}$ with $i=2p+1$, $j=2s+1$ be equal to
$\rho_{ij}$. Then the $n$$\times$$n$ matrix reads
\[
R^{(1)}_{o\,ps}=\left(\sum_{k=1}^{n}\rho^{(1)}_{2k+1,2k+1}\right)^{-1}\rho^{(1)}_{2p+1,2s+1};
\]
it has the properties
$\left(R^{(1)}_{o\,ps}\right)^{\dagger}=R^{(1)}_{o\,ps},$
$\textrm{Tr}(R^{(1)}_{o})=1$ and $R^{(1)}_{o}\geq0$. Thus, we
constructed the positive map $\rho^{(1)}\rightarrow R^{(1)}_{o}$. An
analogous construction can be done by taking the odd matrix elements
equal to zero, whose map can be denoted by $\rho^{(1)}\rightarrow
R^{(1)}_{e}$. If the $N$$\times$$N$ matrix $\rho^{(1)}$ is large
enough, one has the equalities of the von Neumann and linear
entropies,
\[
-\textrm{Tr}\,(\rho^{(1)}\ln\rho^{(1)})\simeq-\textrm{Tr}\,(R^{(1)}_{o}\ln
R^{(1)}_{o})\simeq-\textrm{Tr}\,(R^{(1)}_{e}\ln R^{(1)}_{e}).
\]
The same procedure can be applied to the density matrix of the
bipartite system $\rho^{(12)}$. We only need to change the indices
$i,\,j$ by $\alpha$ and $k,\,l$ by $\beta$, where
$\alpha,\,\beta=1,\,\dots,\,N^{2}$, and consider the odd or even
matrix elements as zeros, and so on.

The entanglement properties of the state $\rho(x,y,x',y')$ are again
preserved for the matrices $R^{(12)}_{o}$ and $R^{(12)}_{e}$. One
can check that global criteria as the logarithmic negativity for the
discretized matrices reflect the phenomenon of entanglement.

\subsection{Cut maps of density matrices}

The discussed example of density $N$$\times$$N$ matrices (with even
number $N=2n$) mapped on smaller density $n$$\times$$n$ matrices is
a particular case of specific nonlinear positive maps. Below we
describe such maps, which we call cut maps; they are constructed as
follows.

First, we consider an arbitrary $N$$\times$$N$ matrix $R$ (not only
the density matrix) with matrix elements $R_{jk}$,
$j,k=1,2,\ldots,N.$ Then, we construct $N$$\times$$N$ matrix
$R'_{jk}$, where arbitrary $k$th columns and rows ($k=1,2,\ldots,m$,
$m<N$) are replaced by columns and rows with zero matrix elements.
The new matrix $R''$ reads
\begin{equation}\label{1}
R''=\frac{R'}{\mbox{Tr}\,R'},
\end{equation}
and we assume that $\mbox{Tr}\,R'\neq 0$.

For example, by this prescription, the 3$\times$3 matrix
\begin{equation*}
R_{jk}=\left(\begin{array}{ccc}
R_{11}&R_{12}&R_{13}\\
R_{21}&R_{22}&R_{23}\\
R_{31}&R_{32}&R_{33}\end{array}\right) 
\end{equation*}
yields the matrices
\begin{eqnarray}\label{2}
R''_{1}&=&\frac{1}{R_{22}+R_{33}}\left(\begin{array}{ccc}
0&0&0\\
0&R_{22}&R_{23}\\
0&R_{32}&R_{33}\end{array}\right),\nonumber \\
R''_{2}&=&\frac{1}{R_{11}+R_{33}} \, \left(\begin{array}{ccc}
R_{11}&0&R_{13}\\
0&0&0\\
R_{31}&0&R_{33}\end{array}\right),\nonumber\\
R''_{3}&=&\frac{1}{R_{11}+R_{22}} \,\left(\begin{array}{ccc}
R_{11}&R_{12}&0\\
R_{21}&R_{22}&0\\
0&0&0\end{array}\right),
\end{eqnarray}
with five zero matrix elements. Three other 3$\times$3 matrices,
which we can obtain by replacing two columns and rows with zero
matrix elements with applying the renormalization tool, are
\begin{equation}\label{3}
R''_4=\left(\begin{array}{ccc}
1&0&0\\
0&0&0\\
0&0&0\\\end{array}\right),\ R''_5=\left(\begin{array}{ccc}
0&0&0\\
0&1&0\\
0&0&0\\\end{array}\right),\ R''_6=\left(\begin{array}{ccc}
0&0&0\\
0&0&0\\
0&0&1\\\end{array}\right).
\end{equation}
We can consider this prescription as a map of 3$\times$3 matrix $R$ onto
2$\times$2 matrices just by removing all zero columns and rows in
matrices~(\ref{2}), i.e., maps of 3$\times$3 matrix $R$ onto three matrices
\begin{eqnarray}\label{4}
\widetilde R''_{1}&=&\frac{1}{R_{22}+R_{33}} \, \left(\begin{array}{cc}
R_{22}&R_{23}\\
R_{32}&R_{33}\end{array}\right),\nonumber \\
\widetilde R''_{2}&=&\frac{1}{R_{11}+R_{33}} \, \left(\begin{array}{cc}
R_{11}&R_{13}\\
R_{31}&R_{33}\end{array}\right),\nonumber\\
\widetilde R''_{3}&=&\frac{1}{R_{11}+R_{22}} \, \left(\begin{array}{cc}
R_{11}&R_{12}\\
R_{21}&R_{22}\end{array}\right).
\end{eqnarray}
Due to this procedure, matrices~(\ref{3}) convert just to the identity
1$\times$1 matrix.

In fact, the procedure we employed is equivalent to cutting some
$k$th columns and rows ($k<N$) in the initial matrix $R$ accompanied
by the normalization (dividing by the trace of the matrix obtained).

The map under discussion is a nonlinear map of the initial
$N$$\times$$N$ matrix on the $n$$\times$$n$ matrices, where $n$ is
an arbitrary integer, such that $n<N$. This map is such that in
cases, where the matrix $R$ has the properties of the density
matrix, i.e., $R\equiv\rho$, with $\rho^\dagger=\rho$, Tr$\,\rho=1$,
and $\rho\geq 0$, the cut map preserves all of them.

For example, if we consider the 3$\times$3 matrix
$R_{jk}=\rho_{jk}$, where $\rho_{jk}$ is the density matrix of the
qutrit state (spin $j=1$ state), we obtain the density matrices of
the qubit state (spin $j=1/2$ state), i.e., $\rho\to\rho_1$,
$\rho\to\rho_2$ and $\rho\to\rho_3$, where
\begin{eqnarray}\label{5}
\rho_{1}&=&\frac{1}{\rho_{22}+\rho_{33}} \, \left(\begin{array}{cc}
\rho_{22}&\rho_{23}\\
\rho_{32}&\rho_{33}\end{array}\right),\nonumber \\
\rho_{2}&=&\frac{1}{\rho_{11}+\rho_{33}} \, \left(\begin{array}{cc}
\rho_{11}&\rho_{13}\\
\rho_{31}&\rho_{33}\end{array}\right),\nonumber\\
\rho_{3}&=&\frac{1}{\rho_{11}+\rho_{22}} \, \left(\begin{array}{cc}
\rho_{11}&\rho_{12}\\
\rho_{21}&\rho_{22}\end{array}\right).
\end{eqnarray}
It is worth noting that the hermiticity, normalization, and
nonnegativity conditions of the matrices obtained are obvious. In
fact, the Sylvester criterion of non-negativity for the qubit
density matrices, obtained by using the cut map, is fulfilled since
all the principal minors of the qubit matrices coincide (up to
positive normalization factor) with a part of principal minors of
the initial matrix $\rho$. By induction, this property can be
checked for an arbitrary $N$$\times$$N$ density matrix $\rho$.

One can describe all cut maps for the density operator, acting in the
$N$-dimensional Hilbert space ${\cal H}$, $\hat\rho\to\hat\rho'$ of the system
state by the relation
\begin{equation}\label{6}
\hat\rho'=\frac{\hat P\hat\rho\hat P}{\mbox{Tr}\,\hat P\hat\rho\hat P}\,,
\end{equation}
where $\hat P$ is ($N-m$)-rank projector $\hat P^\dagger=\hat P$, $\hat
P^2=\hat P$, Tr$\,\hat P=N-m$, with $m$ being the integer such that $1\leq
m<N$. Projector $\hat P$ in (\ref{6}) has only diagonal matrix
elements different from zero.

We discussed the cut maps of density matrices since the properties
of the initial matrix $R$ are partially preserved after applying the
map and obtaining the matrix $R''$. Due to this fact, such
phenomenon as the entanglement and other quantum correlation
properties can be studied using the smaller density $n$$\times$$n$
matrix $\rho''$ obtained from the initial density matrix. The
conservation of the properties of the density matrices can be
characterized by using either the difference of the von Neumann
entropies,
\begin{equation}\label{7}
\Delta_S=-\mbox{Tr}\,\rho\ln\rho+\mbox{Tr}\,\rho''\ln\rho'',
\end{equation}
or the difference of the Tsallis entropies of matrices $\rho$ and $\rho''$,
\begin{equation}\label{8}
\Delta_S^{(q)}=\frac{\mbox{Tr}\,\rho^q-\mbox{Tr}\,\rho''^{q}}{1-q}\,.
\end{equation}
Also relative $q$-entropy
\begin{equation}\label{9}
S^{(q)}(\rho\mid\rho'')=\frac{1}{1-q}\mbox{Tr}\left(\rho^q\rho''^{(1-q)}\right)
\end{equation}
and relative von Neumann entropy~\cite{Umegaki}
\begin{equation}\label{10}
S(\rho\mid\rho'')=\mbox{Tr}(\rho\ln\rho)-\mbox{Tr}(\rho\ln\rho'')
\end{equation}
characterize a degree of the preservation of the properties of quantum
correlations after applying the cut map to the initial density matrix $\rho$.

In formulas~(\ref{7})--(\ref{10}), the matrix $\rho''$ is the $N$$\times$$N$
matrix with zero matrix elements in the columns and rows corresponding to the
action of $(N-m)$-rank projector $\hat P$. The nonnegativity of the matrix
$\rho''$ follows from the positivity of the initial matrix $\rho$. In fact,
the nonnegativity of the density operator $\hat\rho$ acting in the Hilbert
space ${\cal H}$ means that for an arbitrary state vector $\mid\psi\rangle$
one has the inequality
\begin{equation}\label{11}
\langle\psi\mid\hat\rho\mid\psi\rangle\geq 0,
\end{equation}
from which follows that $\langle\psi\mid\hat\rho'\mid\psi\rangle\geq 0$ is
valid for an arbitrary vector $\mid\psi\rangle$, since for the vector
$\mid\varphi\rangle=\hat P\mid\psi\rangle$ one has
$\langle\varphi\mid\hat\rho\mid\varphi\rangle\geq 0$.

This proof is coherent with the application of the Sylvester
criterion discussed above to the matrix representation of the
density operator $\hat\rho'$ for the qutrit state.

For continuous variables, the density matrix $\langle\vec
x\mid\hat\rho\mid\vec x'\rangle=\rho(\vec x,\vec x')$ is infinite
dimensional. Typically, the function of variables $\vec x$ and $\vec
x'$ is the continuous function. In view of this fact, the values of
the function in coordinates $\vec x_0,\vec x'_0$ change a little in
points $\vec x_0+d\vec x_0$ and $\vec x'_0+d\vec x'_0$. Thus, the
discretization procedure of the continuous density matrix does not
change the properties of quantum correlations, if there is no
singular behavior of the function derivatives, or if the shifts of
the arguments (steps of the discretization) are not large.

In the case of finite $N$$\times$$N$ matrices, if one cuts uniformly columns
and rows in the matrix $\rho$ using the small shifts between them, the
correlation properties of the obtained matrix $\rho'$ do not change
essentially. This happens if the dependence of the matrix elements of the
initial matrix $\rho_{jk}$ on the indices $j$ and $k$ is smooth enough. In our
study, we address the problems of application of cut maps in such
cases.

As for the case of bipartite-system states, the density
$N$$\times$$N$ matrix $\rho_{jk}$ at $N=nm$ is interpreted as the
matrix with combined indices $j\equiv(\alpha,\beta)$,
$k\equiv(\alpha',\beta')$, $\alpha=1,2,\ldots,n$ and
$\beta=1,2,\ldots,m$. After this, there appears the possibility to
obtain maps to the density matrix of the first subsystem
$\rho_{\alpha,\alpha'}^{(1)}=\sum_{\beta=1}^m\rho_{\alpha\beta,\alpha'\beta}$
and to the density matrix of the second subsystem
$\rho_{\beta,\beta'}^{(2)}=\sum_{\alpha=1}^n\rho_{\alpha\beta,\alpha\beta'}$.

The cut map applied to the initial matrix
$\rho_{jk}\equiv\rho_{\alpha\beta,\alpha'\beta'}$ induces cut maps
of the subsystem density matrices $\rho_{\alpha,\alpha'}^{(1)}$ and
$\rho_{\beta,\beta'}^{(2)}$. The obtained matrices
$\rho'_{\alpha\beta,\alpha'\beta'}$, $\rho'^{(1)}_{\alpha,\alpha'}$
and $\rho'^{(2)}_{\beta,\beta'}$ preserve some information on the
initial density matrices $\rho$, $\rho^{(1)}$ and $\rho^{(2)}$,
among other things the correlation properties of the subsystem
degrees of freedom, including information on entanglement.

In our study, we focused on the properties of initial density
matrices of continuous variables. Obviously, after the first cut
map, we obtained the finite matrix $\rho(j,k)$, with $j$ and $k$
combined indices, from the matrix $\rho(\vec x,\vec x')$. The
following maps are equivalent to the initial cut map with larger
steps of discretization.

\section{Calculation of the entanglement}
To show the procedure of discretization, we apply this method to the
squeezed vacuum state that has the continuous variable bipartite
representation in variables $x,y,x'$ and $y'$. We determine the
dynamics of some entanglement measurements as the von Neumann,
linear and Tsallis entropies and the logarithmic negativity for the
two-mode squeezed vacuum state in the parametric amplifier.

The squeezed vacuum state is a Gaussian two-mode state. It is known
that a two-mode Gaussian state remains the Gaussian one after the
evolution due to a quadratic Hamiltonian. The analytical results for
the linear and von Neumann entropies of any two-mode Gaussian
density matrix can be obtained using the Wigner
function~\cite{agarwal} and the symplectic eigenvalues of the
covariance matrix~\cite{serafini}. The symplectic eigenvalues can
also be used to calculate the logarithmic negativity~\cite{dodonov}.

The physical system is an optical parametric amplifier described by the Hamiltonian
\begin{equation}
H=\hbar \omega_a a^{\dagger} a+\hbar \omega_b b^{\dagger} b -
\hbar \kappa (a^{\dagger} b^{\dagger} e^{-i \omega t}+ab e^{i \omega
t}),
\end{equation}
where $\omega_a$ is the input signal to amplify, $\omega$ is the
pump frequency that provides the energy to amplify the input signal,
$\omega_b$ is the frequency of the idler mode, and $\kappa$ is the
coupling constant between the system two modes. The time evolution
operator can be written as
\begin{eqnarray}
U(t)&=&e^{-i\omega_{a}N_{a}t}e^{-i\omega_{b}N_{b}t}e^{a_{+}(t)a^{\dagger}b^{\dagger}}  e^{a_{0}(t)(a^{\dagger}a+b^{\dagger}b+1)/2}e^{a_{-}(t)ab},
\end{eqnarray}
with the definitions
\begin{eqnarray}
a_{+}(t)=-\frac{e^{-i\Omega t}\sin\nu t}{\cos(\nu t-\gamma)}, \quad
a_{0}(t)=-i\Omega t+2\ln\left(\frac{\cos\gamma}{\cos(\nu
t-\gamma)}\right), \quad
a_{-}(t) =-\frac{\sin\nu t}{\cos(\nu t-\gamma)}\, .
\end{eqnarray}
where $\Omega=\omega-\omega_a-\omega_b$,
$\nu=\sqrt{\Omega^2/4-\kappa^2}$, and $\gamma~=~\textrm{arctanh}
(\Omega / 2 \nu)$.

The two-mode squeezed vacuum state is the result of applying the
squeeze operator $S(\xi)=\exp\{\xi a b- \xi^* a^{\dagger}
b^{\dagger} \}$ ($\xi=r e^{i \phi}$) to the vacuum state $\vert 0,0
\rangle$; it is given by
\begin{equation}
\vert \beta \rangle=\sqrt{1-\vert \beta \vert^2} \sum_{i=0}^{\infty} \beta^i
\vert i, i \rangle \ ,
\end{equation}
with $\beta=-e^{i \phi} \tanh{r}$. The evolution in a parametric
amplifier is also a squeezed vacuum state that can be expressed as
\begin{equation}
\vert \eta (t) \rangle=\sqrt{1-\vert \eta (t) \vert^2} \sum_{i=0}^{\infty}
\eta^i (t) \vert i, i \rangle \ ,
\end{equation}
where
\begin{eqnarray}
\eta (t)&=&\frac{e^{-i\omega t}}{2 \kappa} \Biggl( \frac{4 \kappa^2 \exp \{
-2 \ln(\cos \nu t-i\sin \nu t \tanh \gamma)\}}{\Omega -2 \kappa \coth r+ 2 i
\nu \tan (\nu t+i\gamma)} - 2 i \nu \tan (\nu t +i \gamma)-\Omega \Biggr) \ ,
\label{21}
\end{eqnarray}
with $\gamma=\textrm{arctanh} (\Omega / 2 \nu)$. This state can be expressed as a Gaussian function of the quadratures $x_1$, $x_2$ as follows
\begin{equation}
\psi_{\eta} (x_1 , x_2) =\frac{\omega_b^{1/4}}{\sqrt{\pi }} \sqrt{\frac{1-\vert \eta (t) \vert ^2}{1-\eta^2 (t)}} \exp \left\{ \frac{1}{1-\eta^2} \left( 2 \sqrt{\omega_b} \eta x_1 x_2- \frac{1+\eta^2}{2} (x_1^2+\omega_b x_2^2) \right) \right\} \, ,
\end{equation}
the discretization of this Gaussian function is used to calculate the entanglement between modes.

\subsection{Tsallis entropy}
The Tsallis entropy~\cite{Tsallis} is a generalization of the von Neumann
entropy~\cite{vonNeumann-entropy}. It is defined by
\begin{equation}
S_q=\frac{1}{1-q} (\textrm{Tr} (\rho^q)-1) \, ,
\end{equation}
where $q$ is a positive real number that is related with the
subadditivity of the system. In a bipartite system $X,Y$, the
quantity $S_q(X)+S_q(Y)+(1-q)S_q(X,Y)$ is different from zero, if
there exists correlation between the subsystems, and is zero if
there is no correlation. When the sign of $1-q$ is positive
($0<q<1$), this region is called the super-additive region, while
for $1-q$ negative ($q>1$) it is called the sub-additive region.
When $q=1$, the Tsallis entropy is equal to the von Neumann entropy,
which is an additive quantity. We also note that for $q=2$ the
Tsallis entropy corresponds to the linear entropy.

Starting with the definition of the partial density matrices
\begin{equation}
\rho^{(1)}_{ik}=\sum_{j} \rho_{ij, kj}\, ,\quad \rho^{(2)}_{jl}=\sum_{i}
\rho_{ij, il} \, ,
\end{equation}
we calculate the eigenvalues of either $\rho^{(1)}_{ik}$ or
$\rho^{(2)}_{jl}$, which are denoted by the
sets $\{e^{(1)} \}$ and $\{e^{(2)} \}$, respectively.

The Tsallis entropy for the reduced system in the discrete scheme
of the density matrix is
\begin{equation}
S^{(i)}_q=\frac{1}{1-q} \left(\sum_k e^{(i)q}_k-1 \right) \, ,
\end{equation}
with $i=1,2$.

The density matrix of the squeezed vacuum state in the parametric
amplifier reads
\[
\rho=(1-\vert \eta \vert^2) \sum_{n,m=0}^{\infty} \eta^n \eta^{*m} \vert n,n
\rangle \langle m,m \vert \, .
\]
Making the partial trace over the second mode, one has
\[
\rho^{(1)}=(1-\vert \eta \vert^2) \sum_{i=0}^{\infty} \vert \eta \vert^{2 i}
\vert i \rangle \langle i \vert \ ,
\]
so one obtains the following eigenvalues for the partial density matrix
\begin{equation}
e^{(1)}_i=(1-\vert \eta \vert^2) \vert \eta \vert^{2 i} \ .
\label{par_eig}
\end{equation}
Using the expression for the eigenvalues of the partial density
matrix in Eq.~(\ref{par_eig}), one has the following expression for
the Tsallis entropy
\begin{equation}
S_q =\frac{1}{1-q} \left( \frac{(1-\vert \eta \vert^2)^q}{1-\vert \eta
\vert^{2q} }-1 \right) \ .
\end{equation}
In Fig.~1, the evolution of the Tsallis entropy ($q=5$) for the
squeezed vacuum state is presented. The entropy exhibits a periodic
behavior with minima at times equal to $t_{\rm min}=m
\, \pi / \nu$ and maxima at times $t_{\rm max}= (2 m +1) \pi
/(2 \nu)$, with $m=0,1,2,\ldots$. The difference of the plots for
the analytical solution and for the cut maps with $n=32$ and $n=16$
are almost negligible.  For $n=8$, there is a region where both
solutions overlap, but the difference grows around the times with
maxima of entropies.  For $n=4$, one can also see that the
approximation is not appropriate, in spite of the fact that some
information on the minimum value of the entanglement and the
periodicity of the curve still remains.

\begin{figure}
\centering
\includegraphics[scale=0.35]{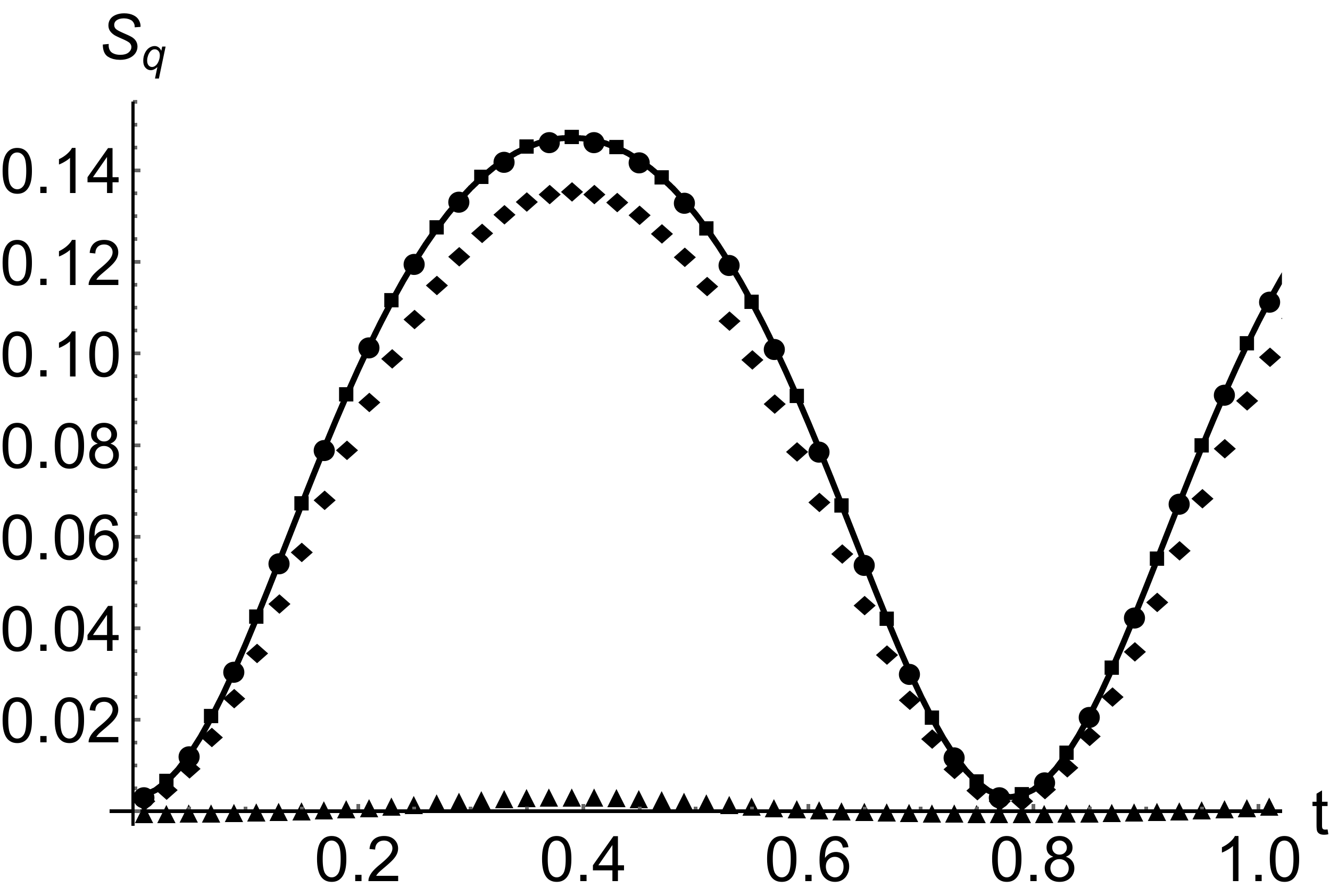}
\caption{Tsallis entropy ($q=5$) as a function of time for the squeezed vacuum
state ($\beta=1/20$) in the parametric amplifier with $\omega_a=1$,
$\omega_b=3$, $\omega=5$, $\kappa=2$, and $\nu=\sqrt{65}/2$. The solid line represents the analytical results, the discrete series denote the numerical results as follows ( circles $n=32$, squares $n=16$, rhombus $n=8$, and triangles $n=4$)}

\label{fig:tsallis_comp}
\end{figure}

\subsection{von Neumann and linear entropies}
The von Neumann and linear entropies can also be obtained through
the discretization procedure. Making use of the eigenvalues of the
partial density matrices, we can write the von Neumann entropy as
follows:
\begin{equation}
S_{\rm VN}=-\sum_{i} e^{(1)}_i \ln e^{(1)}_i=-\sum_{i} e^{(2)}_i \ln
e^{(2)}_i \, .
\end{equation}
The linear entropy can also be calculated similarly,
\begin{equation}
S_L=1-\sum_{i} e^{(1)2}_i=1-\sum_{i} e^{(2)2}_i \, .
\end{equation}
The evolution of these entropies can be evaluated for the squeezed
vacuum state in the parametric amplifier. Using the eigenvalues of
the partial density matrix in Eq.~(\ref{par_eig}) it can be shown
that the linear entropy is given by
\begin{equation}
S_{L}(t)=\frac{2 \left|\eta (t)\right|^{2}}{1+\left|\eta (t) \right|^{2}}\ ,
\label{27}
\end{equation}
with $\eta$ given by Eq.~(\ref{21}).

\begin{figure}
\centering
\includegraphics[scale=0.35]{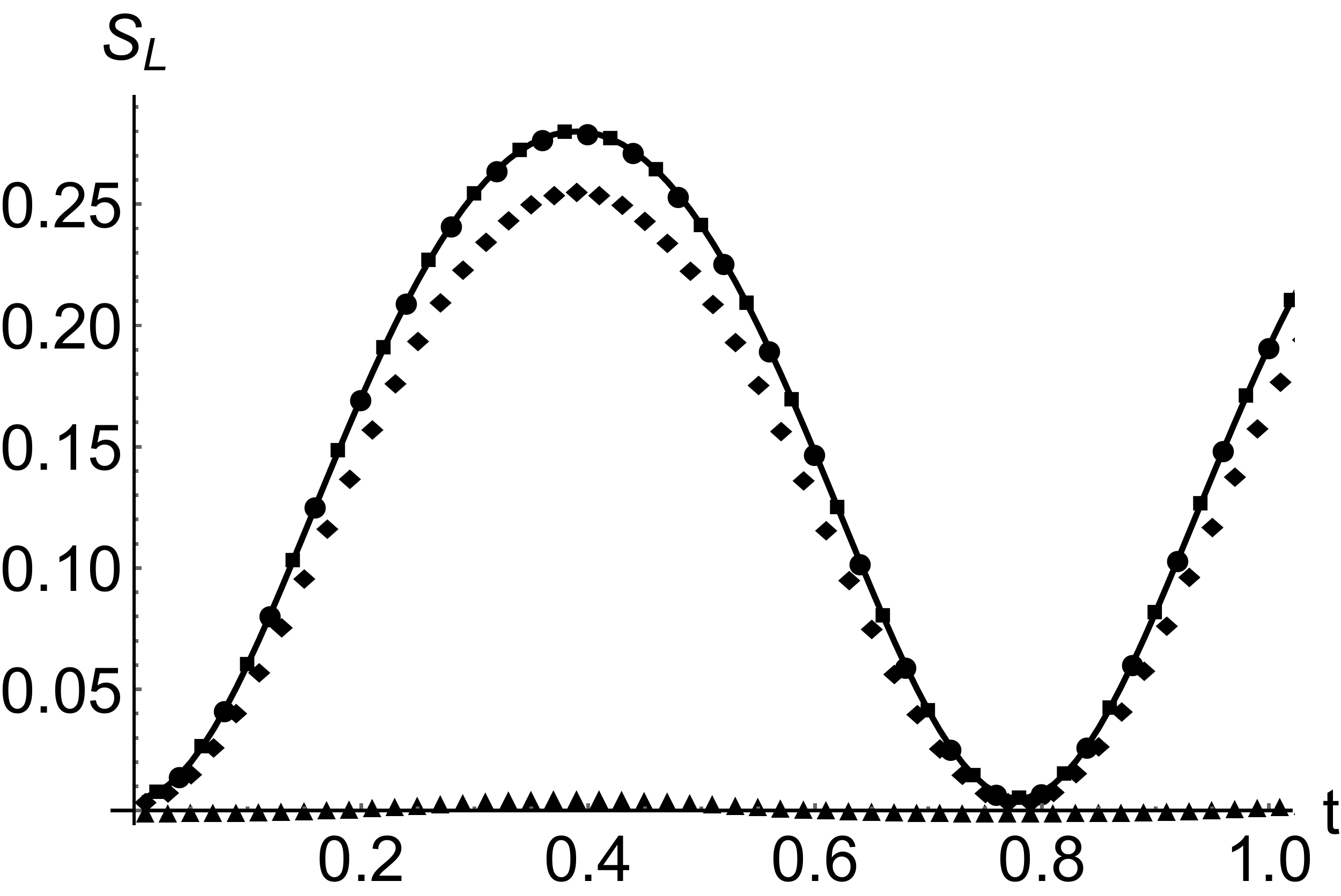}
\caption{Linear entropy as a function of time for the squeezed vacuum state
($\beta=1/20$) in the parametric amplifier with $\omega_a=1$,
$\omega_b=3$, $\omega=5$, $\kappa=2$, and $\nu=\sqrt{65}/2$. The discrete series denote the same cases as fig.~\ref{fig:tsallis_comp}} 
\label{fig:lin_comp}
\end{figure}

Similarly, for the von Neumann entropy we obtain
\begin{equation}
S_{VN}(t)=-\ln\left(1-\left|\eta (t)\right|^{2}\right)-\frac{\left|
\eta (t)\right|^{2}\ln\left(\left|\eta (t)\right|^{2}\right)}{1-\left|
\eta (t)\right|^{2}}\ .
\label{28}
\end{equation}
\begin{figure}
\centering
\includegraphics[scale=0.35]{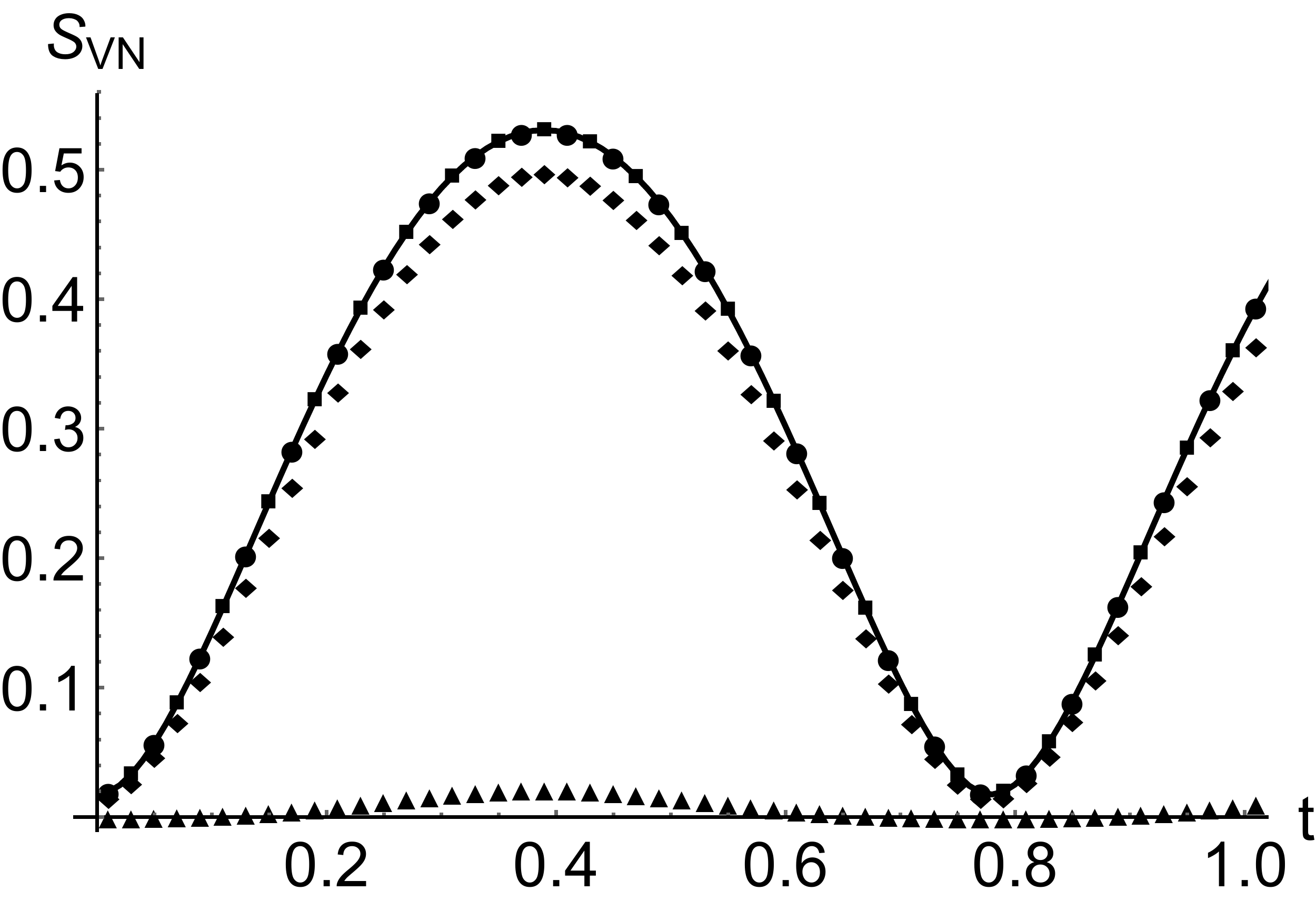}
\caption{von Neumann entropy as a function of time for the squeezed vacuum state
($\beta=1/20$) in the parametric amplifier with $\omega_a=1$,
$\omega_b=3$, $\omega=5$, $\kappa=2$, and $\nu=\sqrt{65}/2$. The discrete series denote the same cases as fig.~\ref{fig:tsallis_comp}}
\label{fig:neumann_comp}
\end{figure}
The dynamics of the linear and von Neumann entropies are plotted in
Figs. 2 and 3, which have the same properties as the Tsallis
entropy. They are periodic ($T=\pi / \nu$), and the cut maps with
$n=32$ and $n=16$ have a remarkable agreement with the analytic
results.  The cut map with $n=8$ differs in the vicinity of the
maximum values region and the cut map with $n=4$ keeps only some
general information on the entropies.

\subsection{Logarithmic negativity}
The logarithmic negativity criterion is an application of the
Peres--Hodorecki criterion~\cite{Peres,Horod} to the density matrix.
This criterion establishes that the partial transposition of a
separable density matrix $\rho=\sum_i p_i \rho^{(1)}_i \otimes
\rho^{(2)}_i$ satisfies the same properties as the original density
matrix, i.e., is hermitian, positively semi-definite and, when there
exists entanglement between the modes, the partial transpose of the
density matrix exhibits negative eigenvalues. Also the sum of the
absolute value of the negative eigenvalues (called negativity)
$\sum_i\vert e^{(-)}_i \vert$ grows, if the system is more
entangled. The logarithmic negativity reads
\begin{equation}
LN=\log_2 \left(2 \sum_i \vert e^{(-)}_i \vert+1 \right) \, ,
\end{equation}
where the set $\{ e^{(-)}_i \} $ represents the negative eigenvalues
of the partial transpose matrix $\rho^{(1)\rm
pt}_{ijkl}~=~\rho_{kjil}$ or $\rho^{(2)\rm pt}_{ijkl}=\rho_{ilkj}$

The partial transpose of the density matrix for the squeezed vacuum state in
the parametric amplifier can be obtained through the expression
\begin{equation}
\rho^{(2)\rm pt}=(1-\vert \eta (t) \vert ^2) \sum_{n,m=0}^{\infty} \eta^n \eta^{*m}
\vert n,m \rangle \langle m,n \vert \ ,
\end{equation}
which has the following states as eigenvectors:
\begin{eqnarray*}
\vert \Phi_0 \rangle&=& \vert r,r \rangle \ , \nonumber \\
\vert \Phi_{\pm} \rangle&=&
\frac{1}{\sqrt{\eta^r \eta^{*s}+\eta^s \eta^{*r}}}\{ \sqrt{\eta^r \eta^{*s}}
\vert r,s \rangle \pm \sqrt{\eta^s \eta^{*r}} \vert s,r \rangle\} \ ,
\end{eqnarray*}
with $r\neq s $. These states have eigenvalues
\[
(1-\vert \eta \vert^2) \vert \eta \vert^{2r}, \ \pm (1-\vert \eta \vert^2)
\vert \eta \vert^{r+s} \ ,
\]
respectively. Thus, the negativity reads
\[
\frac{\vert \eta \vert }{1-\vert \eta \vert } \, ,
\]
and the logarithmic negativity is
\begin{equation}
{\rm LN}=\log_2 \left( \frac{1 + \vert \eta \vert }{1-\vert \eta
\vert }
\right) \, .
\end{equation}
\begin{figure}
\centering
\includegraphics[scale=0.35]{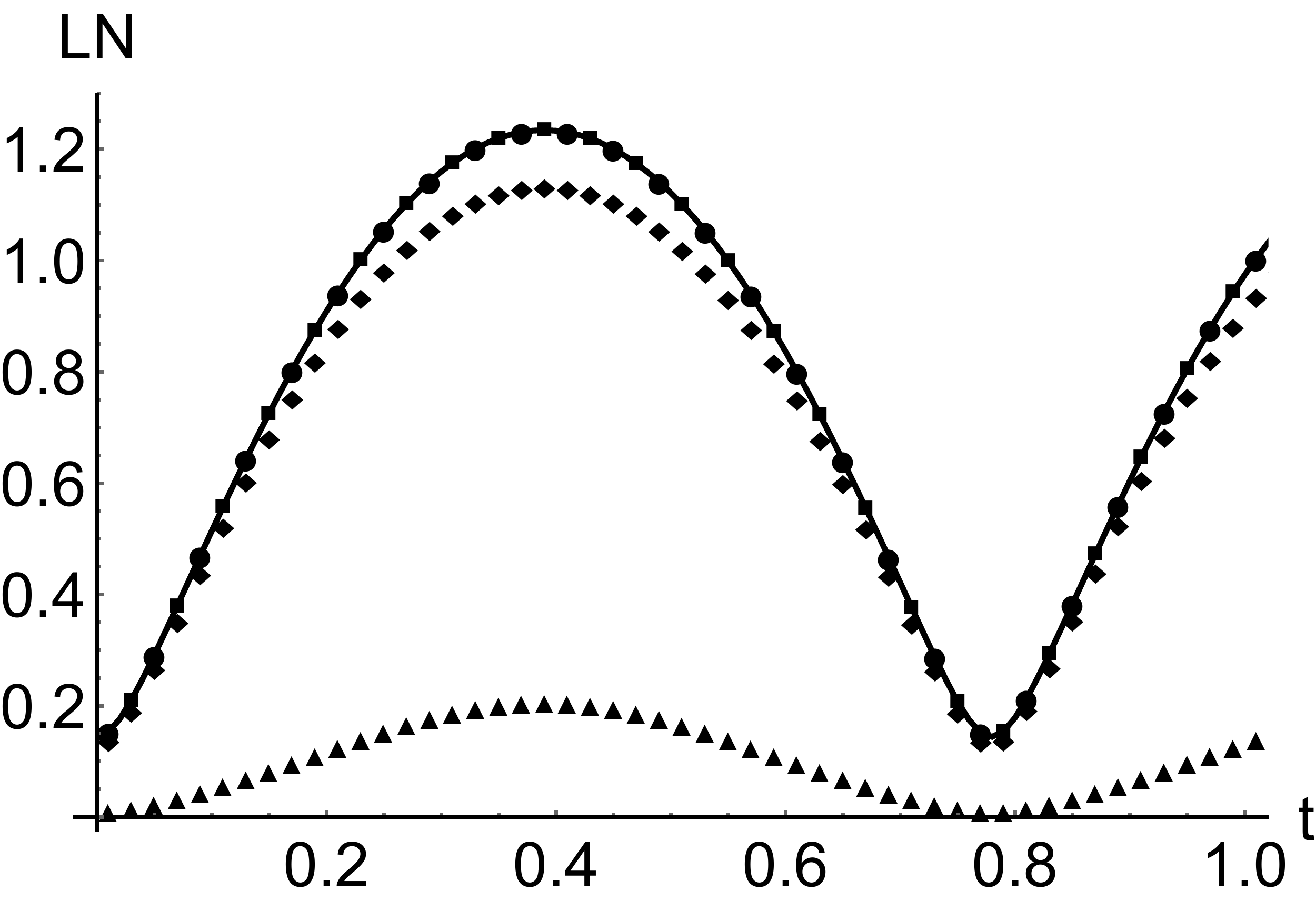}
\caption{Logarithmic negativity as a function of time for the squeezed
vacuum state ($\beta=1/20$) in the parametric amplifier with
$\omega_a=1$, $\omega_b=3$, $\omega=5$, $\kappa=2$, and
$\nu=\sqrt{65}/2$.The solid line represents the analytical results, the discrete series denote the numerical results for a partition of $n$ elements as follows ( circles $n=1024$, squares $n=512$, rhombus $n=256$, and triangles $n=128$)}
\label{fig:nega_comp}
\end{figure}
In Fig.~4, we show the logarithmic negativity as a function of time.
To calculate these quantities, we use a cut map of the bipartite
partial transpose density matrix with $n=1024,512,256,128$. Note
that these values are two orders of magnitude larger, and thus the
procedure is computationally more complex. The differences between
the analytic and numerical results are negligible for the cut maps
with $n=1024$ and $n=512$. For the cut map with $n=256$, one finds
larger discrepancies than the previous maps around the maxima of the
analytic curve. For the cut map with $n=128$, some information on
the logarithmic negativity is preserved, as it happens for the other
entanglement calculations.

\section{Other observables}
The study of other observables obtained through the established
nonlinear map can be carried out similarly.  As an example, we
consider the covariance matrix $\boldsymbol{\sigma}(t)$, which for
the parametric amplifier is defined by
\begin{eqnarray}
\boldsymbol{\sigma}(t)&=&\left( \begin{array}{cccc}
\sigma_{p_1 p_1} & \sigma_{p_1 p_2} & \sigma_{p_1 q_1} & \sigma_{p_1 q_2} \\
\sigma_{p_2 p_1} & \sigma_{p_2 p_2} & \sigma_{p_2 q_1} & \sigma_{p_2 q_2} \\
\sigma_{q_1 p_1} & \sigma_{q_1 p_2} & \sigma_{q_1 q_1} & \sigma_{q_1 q_2} \\
\sigma_{q_2 p_1} & \sigma_{q_2 p_2} & \sigma_{q_2 q_1} & \sigma_{q_2 q_2}
\end{array} \right) \, ,
\end{eqnarray}
with $\sigma_{x y}=\frac{1}{2} \langle \{ x, y\} \rangle-\langle x \rangle
\langle y \rangle$. Using the discretized form of the density matrix, we calculate
different covariances as follows:
\begin{eqnarray}
\sigma_{p_1 p_2}&=&- \sum_{i=1}^{n-1} \sum_{j=1}^{n-1}
\rho_{i+1,j+1,i+1,j+1} 
+  \sum_{i=1}^{n-1}\sum_{j=1}^{n}
\rho_{i+1,j,i+1,j} 
+\sum_{i=1}^{n}\sum_{j=1}^{n-1}\rho_{i,j+1,i,j+1} \nonumber\\
&-&\sum_{i=1}^{n}\sum_{j=1}^{n} \rho_{i,j,i,j}  \, ,  \\
\sigma_{p_k p_k} &=&-\sum_{i=1}^{n-1} \rho^{(k)}_{i,i+1}+2
\sum_{i=2}^{n} \rho^{(k)}_{i,i-1} -\sum_{i=1}^n \rho^{(k)}_{i,i} \, ,
\end{eqnarray}
with $k=1,2$,
\begin{eqnarray}
\sigma_{q_k q_k}&=&\sum_{i=1}^{n} \rho^{(k)}_{i,i} z^2
(\Delta z)^2 
-\left(\sum_{i=1}^{n} \rho^{(k)}_{i,i} z
(\Delta z)^2\right)^2 \, ,
\end{eqnarray}
where for the first mode $z=x$, while for the second one $z=y$. Also
\begin{eqnarray}
\sigma_{q_k p_k}&=&-\frac{i}{2}\Bigl(\sum_{i=1}^{n-1}\rho^{(k)}_{i+1,i}
z-2\sum_{i=1}^{n} \rho^{(k)}_{i,i} z +\sum_{i=1}^{n-1}
\rho^{(k)}_{i+1,i}z\Bigr) \Delta z \, .
\end{eqnarray}

\begin{figure}
\centering
\includegraphics[scale=0.35]{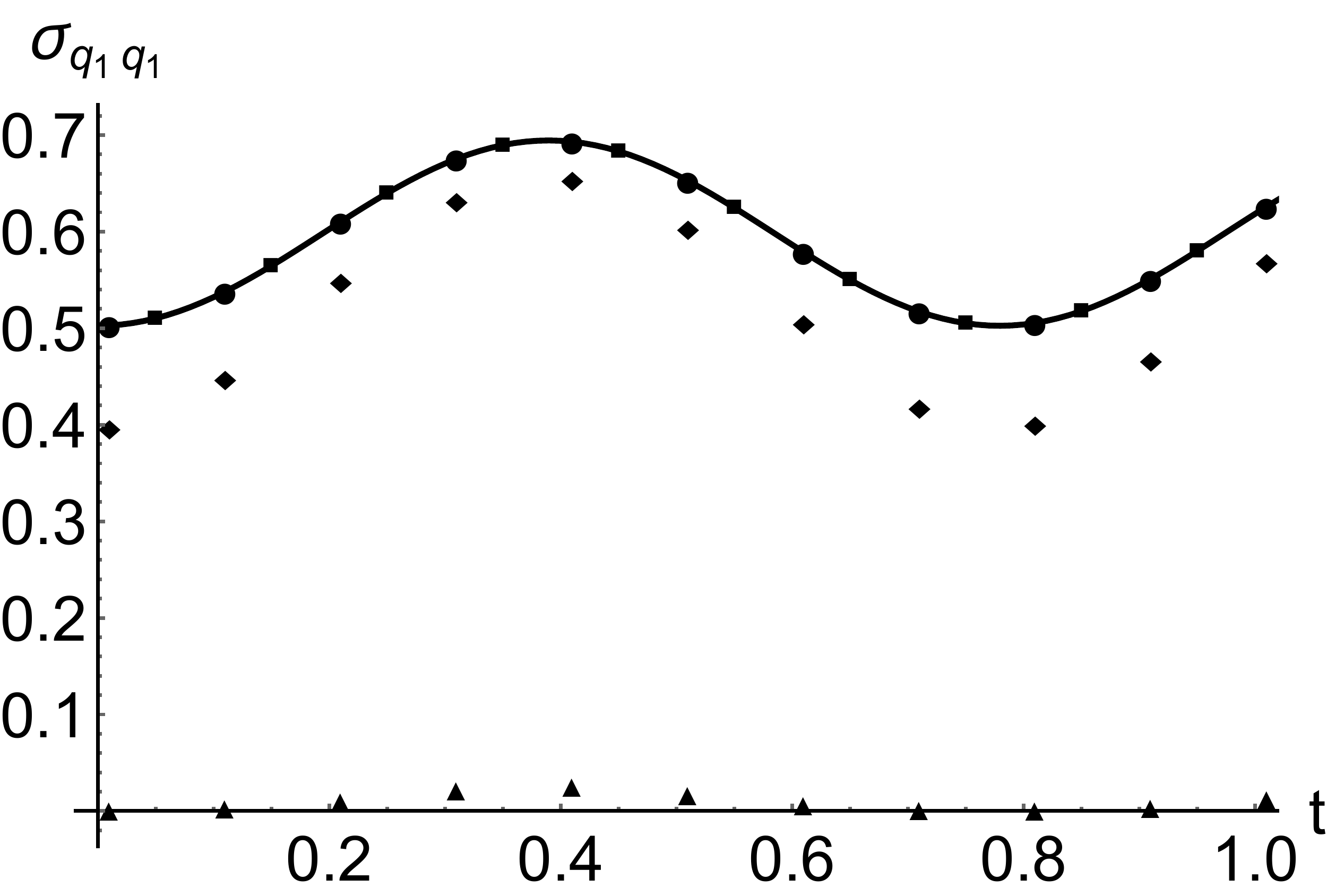}
\caption{The covariance $\sigma_{q_1 q_1}$ as a function of time for the squeezed vacuum state
($\beta=1/20$) in the parametric amplifier with $\omega_a=1$,
$\omega_b=3$, $\omega=5$, $\kappa=2$, and $\nu=\sqrt{65}/2$. The discrete series denote the same cases as fig.~\ref{fig:tsallis_comp}}
\label{fig:sqq}
\end{figure}
\begin{figure}
\centering
\includegraphics[scale=0.35]{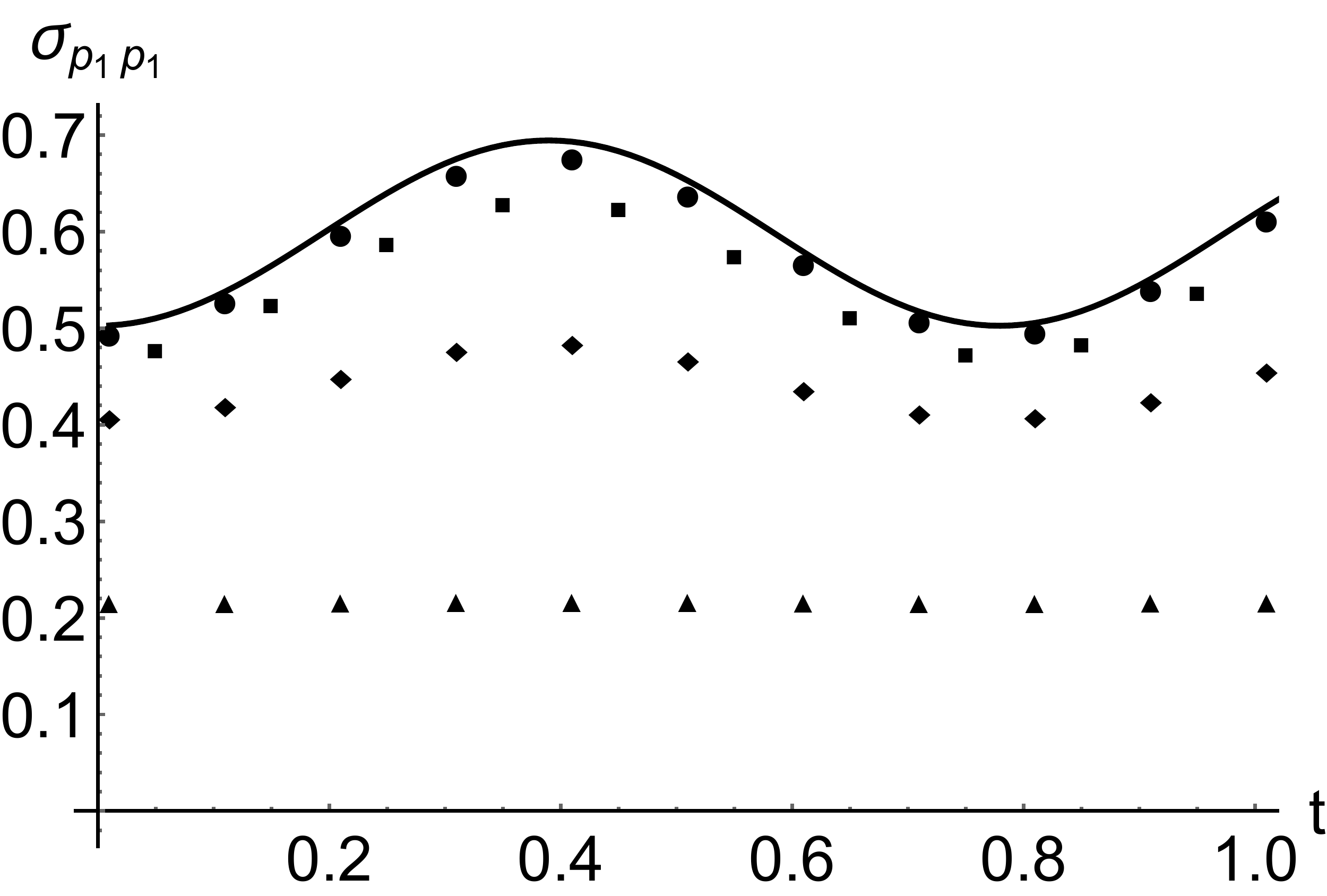}
\caption{The covariance $\sigma_{p_1 p_1}$ as a function of time for the
squeezed vacuum state ($\beta=1/20$) in the parametric amplifier
with $\omega_a=1$, $\omega_b=3$, $\omega=5$, $\kappa=2$, and
$\nu=\sqrt{65}/2$. The discrete series denote the same cases as fig.~\ref{fig:tsallis_comp}}
\label{fig:spp}
\end{figure}The covariance matrix elements $\sigma_{q_1 q_1}$ and $\sigma_{p_1
p_1}$ as functions of time are exhibited in Figs.~\ref{fig:sqq} and
\ref{fig:spp}; we see that the cut maps, showing good agreement
with the analytic results for the entanglement, also display a good
agreement in the covariances matrix elements. In the case of
$\sigma_{p_1 p_1}$, the numerical results are worse than the results
for $\sigma_{q_1 q_1}$, because for the quadrature $p$ it is
necessary to approximate the first derivative (and also the second
derivative) of the density matrix by the expression $\langle x, y
\vert p \rho \vert x', y' \rangle \simeq -i (\rho(x,y+\Delta y; x',
y')-\rho(x,y; x', y'))/\Delta y$. This approximation has an
associated error of $(\Delta y)^2$, which can be noticed in
Fig.~\ref{fig:spp}.

\section{Conclusions}

In this work, we proposed the discretization of the density matrices
of continuous variables as nonlinear positive maps. The resulting
discrete density matrix can be used to calculate the quantum
properties of the system, such as entanglement between the modes.
Specifically, the two-mode entanglement measures used in this work
are the Tsallis, von Neumann and linear entropies and the
logarithmic negativity. This procedure is demonstrated using the
two-mode squeezed vacuum state evolving in the parametric amplifier.
The squeezed vacuum state presents a periodic entanglement with
period $T=\pi /\nu$.

The entanglement measures based on Tsallis entropy, von Neumann
entropy and linear entropy are obtained using the discretization
procedure for the reduced density matrix with different dimensions
$n=32,16,8,4$. The numerical values of entropies for the procedure
using discretization with $n=32$ and $n=16$ are almost equal to the
analytic expressions of those quantities; in the case of $n=8$, the
error between the numerical and analytic results is larger than in
the previous case. In the case of $n=4$, the results are completely
different, although the periodic behavior remains. The logarithmic
negativity, being a two-mode measure of the entanglement, was
obtained using a larger dimension than with the entropies, which can
be obtained using only one mode. In this case, the two-mode density
matrix was discretized using the different dimensions
$n=1024,512,256,128$; the analytic and numerical results are
compared giving a very good agreement for $n=1024$ and $n=512$; in
the case of $n=256$, the results are not always equal; and in the
case of $n=128$, the results are completely different although, as
in the case of entropies, the periodic behavior is still present.

We presented the nonlinear positive maps (called cut maps) of the
density $N$$\times$$N$ matrix onto the density $n$$\times$$n$
matrix, with $n<N$. We demonstrated an example of the cut map for
the case of qutrit density matrix. The structure of the map is
obtained following the procedure of mapping the density matrix of
continuous variables onto a finite-dimensional matrix. It is worth
pointing out that the cut positive maps and the discretization
positive maps provide the density matrices of smaller dimensions,
but these matrices almost preserve information on quantum
correlations available in the initial matrices. Further applications
of the map will be considered in a future paper.

\section*{Acknowledgements}
This work was partially supported by CONACyT-M\'exico (under Project
No.~238494) and DGAPA-UNAM (under Project No.~IN110114).

\end{document}